\documentclass[twocolumn,prb,superscriptaddress]{revtex4}
\usepackage{graphicx}
\usepackage{amssymb}
\begin{document}

\title{Effects due to backscattering and pseudogap features in graphene nanoribbons with single vacancies}

\author{I. Deretzis}
\email{ioannis.deretzis@imm.cnr.it}
\affiliation{Scuola Superiore, Universit\`{a} di Catania, I-95123 Catania, Italy}
\affiliation{CNR-IMM, I-95121 Catania, Italy}

\author{G. Fiori}
\affiliation{Dipartimento di Ingegneria dell'Informazione: Elettronica, Informatica, Telecomunicazioni, Universit\`{a} di Pisa, I-56122 Pisa, Italy}

\author{G. Iannaccone}
\affiliation{Dipartimento di Ingegneria dell'Informazione: Elettronica, Informatica, Telecomunicazioni, Universit\`{a} di Pisa, I-56122 Pisa, Italy}

\author{A. La Magna}
\affiliation{CNR-IMM, I-95121 Catania, Italy}
\date{\today}

\begin{abstract}
We present a systematic study of electron backscattering phenomena during conduction for graphene nanoribbons with single-vacancy scatterers and dimensions within the capabilities of modern lithographic techniques. Our analysis builds upon an \textit{ab initio} parameterized semiempirical model that breaks electron-hole symmetry and nonequilibrium Green's function methods for the calculation of the conductance distribution $g$. The underlying mechanism is based on wavefunction localizations and perturbations that in the case of the first $\pi-\pi{}^*$ plateau can give rise to impurity-like pseudogaps with both donor and acceptor characteristics. Confinement and geometry are crucial for the manifestation of such effects. Self-consistent quantum transport calculations characterize vacancies as local charging centers that can induce electrostatic inhomogeneities on the ribbon topology. 
\end{abstract}

\maketitle

Graphene is a stable, two-dimensional, $sp^2$-bonded carbon allotrope system with exceptional electrical, mechanical and optical properties\cite{neto:109,avouris_prov}. Particularity stems from the almost linear dispersion relation in momentum space close to the Fermi level, where valence and conduction bands meet exactly at the charge neutrality point. Provided that a viable method of controllable band-gap engineering can be obtained, graphene can also constitute a valid alternative for post-Si CMOS technology. Graphene's peculiar conduction properties can be highly compromised by the presence of disorder in terms of local scattering centers (e.g. defects\cite{2009PhRvL.102w6805C}) or substrate-induced interference\cite{2009NatMa...8..203E,2009ApPhL..95f3111D}. On the other hand, disorder by means of chemical functionalization has been proposed to enhance device functionality by controlling the opening of exploitable gaps. Dopants can either act directly on the electronic structure\cite{boukhvalov:085413,2009PhRvB..79p5431G}, or induce backscattering effects within the transport process\cite{2009PhRvL.102i6803B}. As a consequence, the study of conduction in disordered graphene systems acquires a double significance: on the one hand it becomes crucial for performance-related characteristics, while on the other it can be used to address band-gap tailoring issues. 

Vacancies can make part of crystalline graphene as production faults of the mechanical or the epitaxial growth process. Their presence has been experimentally verified by transmission electron microscopy\cite{2008NanoL...8.3582M}, whereas structural, electronic and magnetic properties have been theoretically investigated at a full quantum scale\cite{2008PhRvB..77k5109P,2006PhRvL..96d6806C,2000PhRvB..6114089H,2008PhRvB..77s5428P,yazyev:125408}. 
Modeling within the nearest-neighbor tight-binding formalism is straightforward: a local point potential $U\rightarrow \infty$ is introduced on the vacancy site that forbids hoppings to and from neighboring sites\cite{2008PhRvB..77k5109P}. As a result a semi-localized state appears at the Fermi level of the system with a $C_3$ point symmetry and a local spin\cite{2008PhRvB..77k5109P,2008PhRvB..77s5428P}. The positioning of this zero-energy mode is highly related to the electron-hole symmetrical description of the bandstructure under this approach. By considering a next-to-near neighbor in the same type of model, valence and conduction band mirror-symmetry breaks, and the defect state looses its high-symmetry allocation. Accurate \textit{ab initio} calculations show that the exact resonance of this mode is located at energies lower than the charge neutrality point\cite{2006PhRvL..96d6806C}. Pereira \textit{et al.}\cite{2008PhRvB..77k5109P} have argued that the breaking of electron-hole symmetry in the presence of a point defect can induce further alterations in the electronic structure by means of other semi-localized eigenstates, implying that the perturbation induced by the point defect is not constrained to a single energy but it expands within the energy spectrum. Unlikely, a clear allocation of the role of such perturbation with respect to the conduction properties of graphene-based systems still lacks. 

Under this perspective, the objective of the present work is to extensively investigate the importance of wavefunction localization induced by defects for the electrical transport properties of confined graphene. Here we present a systematic investigation of conduction and charging for single-vacancy damaged armchair and zigzag graphene nanoribbons (aGNRs and zGNRs respectively) of dimensions within the capabilities of modern lithographic techniques (width up to 4.6 nm). The basis of the transport formalism is a first-principles parameterized semiempirical Hamiltonian that considers atomic interactions on the evaluation of distance-dependent overlap integrals, hence, introducing further neighbor interactions in a natural way. Computational results show that there is a clear relationship between the position of the vacancy, the resonance of the defect-states and the local eigenvector value of the corresponding unperturbated system. Coupling between such features can give rise to resonant backscattering phenomena during the conduction process that in the case of the first $\pi-\pi{}^*$ plateau are associated with the opening of pseudogaps, similar to ones obtained for $p$-type impurities\cite{2009PhRvL.102i6803B}. This type of phenomenon emerges for all ribbons within this study. Notwithstanding this, a vast complexity characterizes the results from case to case, since the eigenvector nature of the problem imposes a strict relationship with respect to the system geometry in terms of confinement, position and symmetry. Moreover self-consistent Schr\"{o}dinger/Poisson calculations show that vacancies become charging centers in a nonequilibrium process and induce long-range carrier-density inhomogeneities on the ribbon topology.

We consider the convention of Ref. \citenum{2006PhRvL..97u6803S} to classify aGNRs (zGNRs) on the number of the dimer lines $N_a$ (zigzag chains $N_z$) across the ribbon width. Electronic Hamiltonians are formed on the basis of the extended H\"{u}ckel theory\cite{kienle:043714}, using a double-$\zeta$ Slater basis set that derives by fitting the bandstructure of bulk graphene according to density functional calculations (see Ref. \citenum{kienle:043714} for a detailed description of the parameterization). Energy eigenstates $\epsilon_\alpha$ and their respective eigenfunctions $\Psi_\alpha$ are obtained through a direct diagonalization of the Hamiltonian matrix imposing periodic boundary conditions in supercells up to 8.5 nm long. Quantum transport is studied within the non-equilibrium Green's function formalism (NEGF) coupled to the Landauer-Buttiker approach for the calculation of the conductance $g$\cite{Datta}. Simulations take place within the ballistic regime. Contacts are ideal, i.e. of the same width $N_a$ ($N_z$) as the device without the presence of vacancies. Edges are passivated with single hydrogens.

\begin{figure}
	\centering
		\includegraphics[width=\columnwidth]{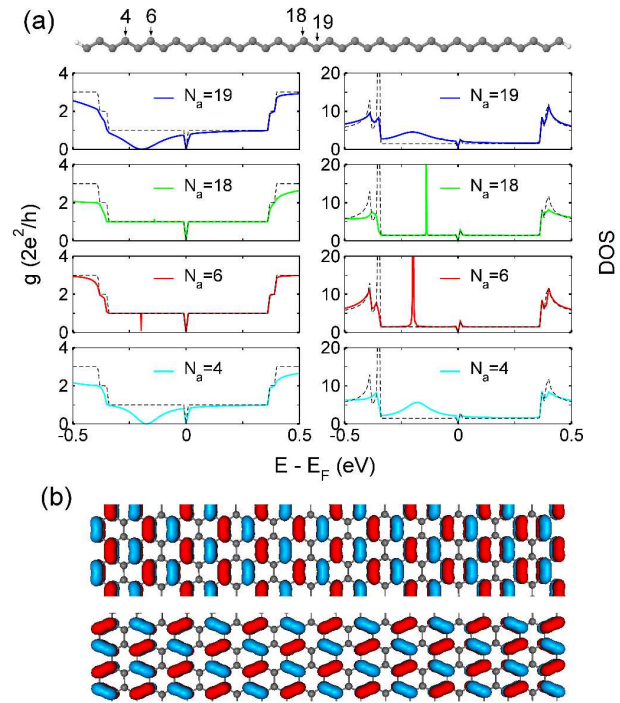}
	\caption{(a) Conductance $g$ and DOS as a function of energy for a 38-aGNR with single vacancies at different positions $N_a$. (b) Projection of the wavefunction corresponding to first state below the charge neutrality level (upper) and the first state above the charge neutrality level (lower) for a non-defected 38-aGNR.}
	\label{fig:figure1}
\end{figure}

We start this study with electronic structure, conductance and density of states (DOS) spectra calculated for a semimetallic 38-aGNR. Here the vacancy is introduced in various positions of a chain transversal to the longitudinal axis of the system (see fig. \ref{fig:figure1}a). Electronic structure calculations show that the universal response of the defect is to give rise to quasilocalized states\cite{2008PhRvB..77k5109P} along nonlocalized ones, that due to electron-hole disparity do not preserve a mirror symmetry with respect to the charge neutrality point. The level of localization for these states can vary from strong to weak while their energy resonance highly depends on the position of the defect site. When it comes to conduction, there are two distinct groups of behaviors obtained on the basis of the location of the vacancy sites. For sites of the first group, significant conductance dips and pseudogap features are present within the valence band of the first $\pi-\pi{}^*$ plateau of the system. The correspondent DOS interestingly shows a smooth bell-like region which expands throughout the conductance dip zone. The presence of a non-negligible DOS where a conduction gap takes place is strongly correlated with resonant backscattering phenomena during the transport process. Such feature is strikingly similar to analogous effects calculated for $p$-type (nominally boron) impurities\cite{2009PhRvL.102i6803B}. On the other hand, second group sites practically leave the pure structure's $g=1$ (in units of $2h^2/e$) conductance plateau unaltered. Here the first vacancy state below the charge neutrality point gives rise to a DOS with a sharp peak that decays exponentially and provokes a strongly localized perturbation in the electronic structure. A systematic analysis for the sum of possible vacant sites shows that two thirds of the total sites belong to the first group, while the remaining one third sites conform with the second group. This categorization strictly stands only for the first $\pi-\pi{}^*$ region, since for energies that are more distant from the charge neutrality point, further divergences occur in the conduction properties according to purely geometrical criteria.

\begin{figure}
	\centering
		\includegraphics[width=\columnwidth]{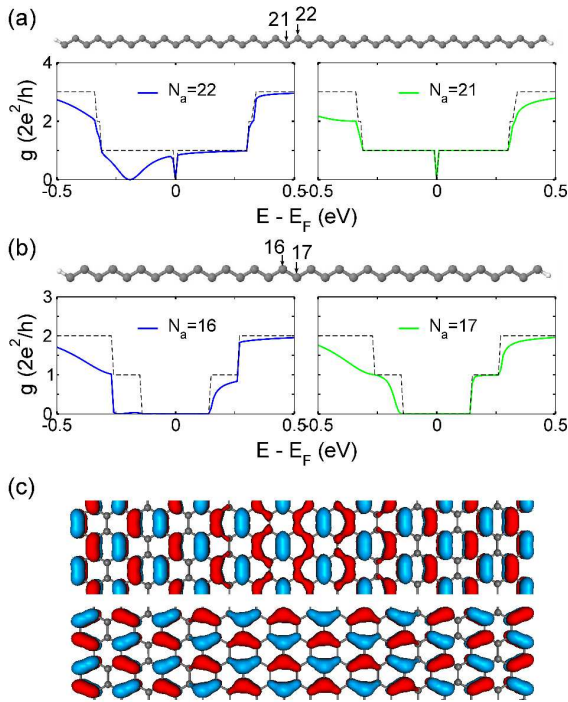}
	\caption{a) Conductance $g$ as a function of energy for a semimetallic 44aGNR with a single vacancy at $N_a=22$ (left) and $N_a=21$ (right). a) Conductance distribution $g$ as a function of energy for a semiconducting 34aGNR with a single vacancy at $N_a=16$ (left) and $N_a=17$ (right). (c) Projection of the wavefunction corresponding to first state below the charge neutrality level (upper) and the first state above the charge neutrality level (lower) for a non-defected 34-aGNR.}
	\label{fig:figure2}
\end{figure}

The $1/3-2/3$ relationship presented above can be understood with a careful examination of the ribbon's wavefunctions in the ideal case, i.e. when no alteration of the atomic structure is present (see fig. \ref{fig:figure1}b). The form of the eigenvectors within the the entire first conductance plateau preserves a $1/3-2/3$ schema where $2/3$ of the sites have a finite eigenvector value while the rest $1/3$ have an extremely small value instead. If the vacancy is introduced in a position corresponding to the $2/3$ group, eigenvector symmetry breaks and defect-mode perturbations spread also to neighboring energies. The sum of perturbative behaviors due to the quasilocalized states within this plateau gives rise to an increased DOS distribution with respect to the ideal case below the charge neutrality point. This expanded perturbation is the reason for electron backscattering during conduction. On the other hand, if the vacancy is introduced in the $1/3$ group sites, the perturbation induced remains localized since the $2/3$ symmetry does not break, and no generalized repercussions are inferred in the conductive capacity of the system. This picture is not only present in the previous aGNR but reflects a general situation for the conduction properties of vacancies in semimetallic nanoribbons with the armchair confinement (e.g. see fig. \ref{fig:figure2}a). A systematic data analysis for aGNRs up to $N_a=44$ dimer lines leads to the following empirical rule (which is supported by analytical calculations of the first plateau wavefunctions for all metallic aGNRs\cite{2007PhRvB..75p5414Z}): for $N_a=3p+2$ dimer lines ($\forall p \in \mathbb{N}$), vacancies at the $N_a=3q$ sites belong to the $1/3$ group ($\forall q \in \mathbb{N}, \leq p$), while the rest make part of the second group. Apart from the semimetallic ribbons, semiconducting aGNRs behave within a similar framework that infers a less pronounced effect on the transport mechanism due to the intrinsic electronic bandgap (see fig. \ref{fig:figure2}b). Unlike though their semimetallic counterparts, wavefunctions of semiconducting aGNRs do not globally preserve the $1/3-2/3$ symmetry within the first conductance plateau (fig. \ref{fig:figure2}c). Notwithstanding this, the mechanism that gives rise to DOS perturbations and therefore to conductance dips remains the same, even if mainly for vacancies towards the center of the ribbons backscattering effects are always present and differences are attenuated.

\begin{figure}
	\centering
		\includegraphics[width=\columnwidth]{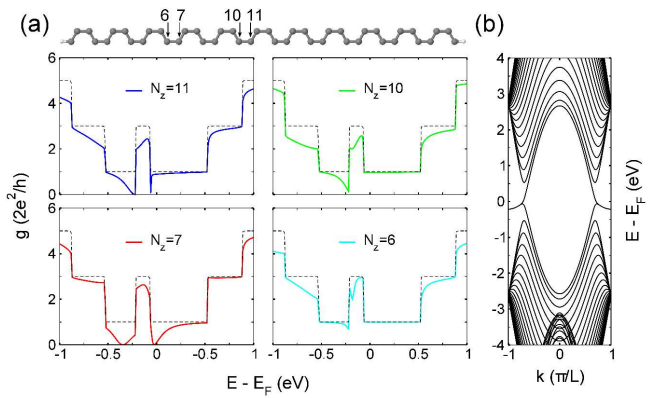}
	\caption{(a) Conductance $g$ as a function of energy for a 22-zGNR with a single vacancy at the $N_z=11$ (up-left), $N_z=10$ (up-right), $N_z=7$ (down-left) and $N_z=6$ (down-right) positions. (b) Bandstructure of an ideal 22-zGNR with periodicity L within the spin-restricted extended H\"{u}ckel model.}
	\label{fig:figure3}
\end{figure}

The case of defected zGNRs is intrinsically more complicated than that of aGNRs due to the presence of edge magnetism (that gives rise to a secondary bandgap\cite{2006PhRvL..97u6803S}) in conjunction with the local spin attributes of vacancies themselves\cite{2008PhRvB..77s5428P}. On the other hand recent \textit{ab initio} calculations have shown that the most stable hydrogen-terminated zig-zag ribbons are not monohydrogenated and loose their magnetic ground state\cite{PhysRevLett.101.096402}. Here we neglect spin interactions and focus on the effect of vacancy-induced pseudogaps within the conduction process, which is complementary to but of a higher order with respect to the intrinsic bandgap (see Ref. \citenum{oeiras:073405} for a discussion on spin-polarized transport). Fig. \ref{fig:figure3} shows representative conductance results for a 22-zGNR with single vacancies at four different internal positions of the ribbon along a chain transversal to the longitudinal axis of the system. The $1/3-2/3$ symmetry seen in the case of aGNRs is not confirmed, reflecting the different form of zGNR wavefunctions. On the other hand symmetry effects with respect to parity, as well as important differences with respect to the positioning of the various defected sites can be obtained. In detail, vacancies at the central region of the zGNR show a $p$-type conduction gap that is greatly compromised by the presence of the $g=3$ plateau area due to band-bending close to the Fermi level near the Brillouin zone boundaries (see Fig. \ref{fig:figure3}b). Moving towards the ribbon's edges, even-$N_z$ sites become completely metallic whereas odd-$N_z$ sites give rise to a simultaneous transport gap in both conduction and valence bands\cite{2009Nanot..20a5201G}. This behavior is similar to impurity backscattering with both donor and acceptor characteristics, whereas analogous effects have been calculated for substitutional boron atoms\cite{2009PhRvL.102i6803B}. The key issue for understanding resonant backscattering phenomena in zGNRs can be traced back in the high local density of states concentration within the $\pi-\pi{}^*$ plateau that is not only limited to edge states, i.e. nonlocalized states in an ideal zGNR are present in this energy zone. The wavefunctions of these states usually maintain a parity symmetry for neighboring sites that can loose balance while moving from the center towards the edges of the ribbon. According to the exact positioning of the vacancy site the perturbation induced can be stronger or weaker while it can affect more than one regions at the entire plateau, giving rise to separate resonant backscattering phenomena during the conduction process. 

\begin{figure}
	\centering
		\includegraphics[width=\columnwidth]{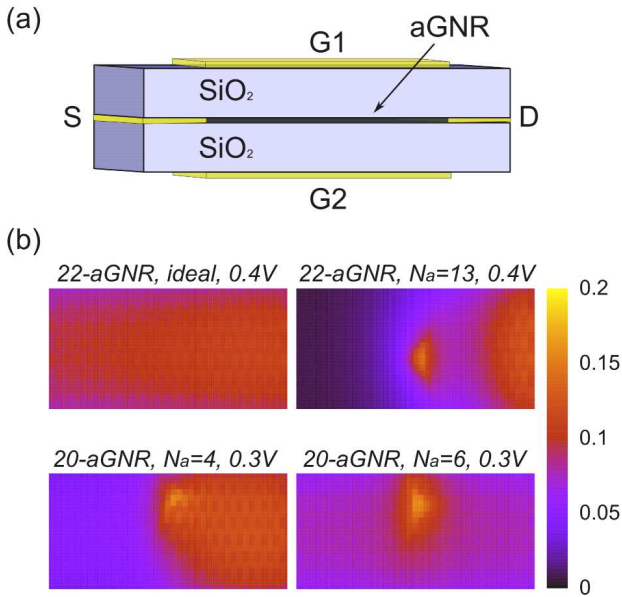}
	\caption{(a) Schematic representation of the simulated 4-terminal system. Front and back gates are isolated by a 1.9 nm thick $SiO_2$ layer with a relative dielectric constant $\kappa = 3.9{}$. Armchair graphene nanoribbons are used as channel materials with length $l=8.41nm$. (b) Maps of the electrostatic potential $\Phi$ (in $V$) along the ribbon surface for an ideal 22-aGNR ($V_{DS}=0.4V$), a 22-aGNR with a single vacancy at $N_a=13,l=4.38nm$ ($V_{DS}=0.4V$), a 20-aGNR with a single vacancy at $N_a=4,l=4.28nm$ ($V_{DS}=0.3V$) and a 20-aGNR with a single vacancy at $N_a=6,l=4.28nm$ ($V_{DS}=0.3V$).}
	\label{fig:figure4}
\end{figure}

For a more realistic description of quantum transport and the evaluation of nonequilibrium aspects of conduction, we have performed self-consistent simulations within the coupled NEGF-Poisson scheme in order to evidence the manifestation of charging effects related to the presence of vacancies in the atomic lattice. To this purpose, we have extended the open-source \textit{NanoTCAD ViDES} code\cite{2007IEDL...28..760F,2006ITED...53.1782F} with extended-H\"{u}ckel functionality and enhanced matrix operations with optimized numerical techniques\cite{Petersen20095020}. We consider the 4-terminal geometry of fig. \ref{fig:figure4}a and 20/22aGNRs with length $l=8.41nm$ as channel materials. Single vacancies are introduced at various positions of the atomic lattice, whereas the relatively short length of these aGNRs allows for a short-range Coulomb scattering study within the ballistic limit. Simulation temperatures are set to $T=300K$. Gate voltages are fixed to zero ($V_{G_1S}=0V, V_{G_2S}=0V$) while drain-to-source bias is gradually raised from $V_{DS}=0V$ to $V_{DS}=0.5V$. Maps of the electrostatic potential $\Phi$ along the ribbon surfaces at the end of the self-consistent process can be seen at fig. \ref{fig:figure4}b. There are two main charging aspects related to the presence of vacancies in these samples: on one hand the immediate region around the defect acquires a charge concentration that manifests as an electrostatic peak on the GNR topology. This feature appears for all types of vacancies discussed in the previous paragraphs. Directionality of such potential is also visible in correspondence to the geometrical orientation of the vacancy. On the other hand though, an important aspect arises in the cases where vacancies are associated with impurity-like backscattering issues. Here the presence of the defect gives rise to a charge-carrier inhomogeneity effect on the GNR topology (e.g. see the lower panel of fig. \ref{fig:figure4}b). Moreover, this process is bias-dependent, i.e. by increasing $V_{DS}$ a more intense electrostatic inhomogeneity pattern can be observed. The presence of carrier inhomogeneities on the graphene surface have been widely attributed to charged impurities, and under this perspective important experimental features related to the measured conductivity of graphene have been justified\cite{2008NatPh...4..377C}. This study  indicates that alternatively, also vacancies can create similar charging disorder whose extend should ideally be calculated for flakes of dimensions within the range of the microscale.    

To conclude, this study has extensively investigated conduction and charging properties of armchair and zigzag graphene nanoribbons with single vacancies. Focus has been put on the underlying backscattering mechanism that proves fundamental for transport-related features like the appearance of pseudogaps within the first conductance plateau. A front-end consequence arising from this mechanism is that vacancies can behave as $p$-type impurities, while additional donor-like behaviors can be observed in the case of zGNRs. It has been argued that the positioning of the defect-states within the eigenspectrum in conjunction with geometrical considerations that shape the system wavefunctions are the origin of the presented phenomena. Moreover self-consistent quantum transport calculations have evidenced that during nonequilibrium, vacancies can induce inhomogeneities in the electrostatic topology of the ribbons, in accordance with similar effects often attributed to the presence of charged impurities. On the basis of such assumptions some key points need to be discussed. From an application point of view the association of vacancies with conduction gaps can have a big practical impact on the engineering of mobility gaps in graphene-based systems, since vacancies are either present in the atomic lattice from the production stage or can be easily obtained e.g. by ion irradiation. Vacancy-concentration has to remain low though in order to avoid predominant inelastic electron-phonon scattering processes and consequently high reductions of the electron mobility\cite{2009PhRvL.102w6805C}. A preferential experimental verification is necessary here. From a methodological point of view an important issue arises with respect to the resonance of the defect states. An inaccurate positioning of these modes in the first conductance plateau can lead to a dislocation of the pseudo-gap resonance, or in some cases, to a complete suppression of such effect. In this sense, second or higher neighbor atomistic models, or impurity-like calibrations of the vacancy site\cite{PRBinpress} seem more appropriate for quantum transport calculations in defected graphene systems.  

\begin{acknowledgments}
G. F. and G. I. would like to acknowledge the EC Seventh Framework Program under Project GRAND (Contract 215752) for partial financial support.
\end{acknowledgments}

\bibliography{VacanciesGraphene}

\end{document}